\documentclass[useAMS,usenatbib]{mn2e}

\usepackage{ulem} 
\usepackage{graphicx}
\usepackage{amsmath}
\usepackage{ulem}
\usepackage{rotating}
\usepackage{amssymb}
\usepackage{url}
\usepackage{hyperref}

\def\lapprox{\hbox{\lower .8ex\hbox{$\,\buildrel < \over\sim\,$}}}
\def\gapprox{\hbox{\lower .8ex\hbox{$\,\buildrel > \over\sim\,$}}}

\title[\text{HST} on NGC\,6752. III. The WDs cooling sequence ]{
  The \textit{HST} Large Programme on NGC\,6752. III. Detection of the Peak of the White Dwarf Luminosity Function\thanks{ 
Based on observations with the NASA/ESA {\it Hubble
Space Telescope}, obtained at the Space Telescope Science Institute,
which is operated by AURA, Inc., under NASA contract NAS 5-26555.
}
}
\author[L.\,R.\,Bedin et al.]{L.\,R.\,Bedin$^{1}$\thanks{E-mail: luigi.bedin@inaf.it}, M.\,Salaris$^{2}$, J.\,Anderson$^{3}$, M.\,Libralato$^{3}$, D.\,Apai$^{4,5}$, D.\,Nardiello$^{6}$,
\newauthor
R.\,M.\,Rich$^{7}$, A.\,Bellini$^{3}$, A.\,Dieball$^{8}$, P.\,Bergeron$^{9}$, A.\,J.\,Burgasser$^{10}$,  A.\,P.\,Milone$^{6}$,
\newauthor
and  A.\,F.\,Marino$^{6}$ ~\\
$^{1}$INAF-Osservatorio Astronomico di Padova, Vicolo dell'Osservatorio 5, I-35122 Padova, Italy\\
$^{2}$Astrophysics Research Institute, Liverpool John Moores University,146 Brownlow Hill, Liverpool L3 5RF, UK\\
$^{3}$Space Telescope Science Institute, 3800 San Martin Drive, Baltimore, MD 21218, USA\\ 
$^{4}$Department of Astronomy and Steward Observatory, The University of Arizona, 933 N. Cherry Avenue, Tucson, AZ 85721, USA\\
$^{5}$Lunar and Planetary Laboratory, The University of Arizona, 1640 E. University Blvd., Tucson, AZ 85721, USA\\
$^{6}$Dipartimento di Fisica e Astronomia ‘Galileo Galilei’, Università di Padova, Vicolo dell’Osservatorio 3, Padova I-35122, Italy\\
$^{7}$Department of Physics and Astronomy, UCLA, 430 Portola Plaza, Box 951547, Los Angeles, CA 90095-1547, USA\\
$^{8}$Argelander Institut f\"ur Astronomie, Helmholtz Institut f\"ur Strahlen-und Kernphysik, University of Bonn, Germany\\
$^{9}$D\'epartement de Physique, Universit\'e de Montr\'eal, C.P.\,6128, Succ.\,Centre-Ville, Montr\'eal, QC\,H3C\,3J7, Canada\\
$^{10}$Center for Astrophysics and Space Science, University of California San Diego, La Jolla, CA 92093, USA
}

\begin{document} 

\date{Accepted 2019 July 12. Received 2019 June 20; in original form 2019 March 28}

\pagerange{\pageref{firstpage}--\pageref{lastpage}} \pubyear{201X}

\maketitle
 
\label{firstpage}

\begin{abstract}
We report on the white dwarf cooling sequence of the old globular
cluster NGC\,6752, which is chemically complex and hosts a blue
horizontal branch.
This is one of the last globular cluster white dwarf (WD) cooling
sequences accessible to imaging by the \textit{Hubble Space Telescope}.
Our photometry and completeness tests show that we have reached the
peak of the luminosity function of the WD cooling sequence, at a
magnitude $m_{\rm F606W}$=29.4$\pm$0.1, which is consistent with a
formal age of $\sim$14\,Gyr. This age is also consistent with the age
from fits to the main-sequence turnoff (13-14\,Gyr), reinforcing our
conclusion that we observe the expected accumulation of white dwarfs
along the cooling sequence.
\end{abstract}


%
\section{Introduction}
\label{introduction}
%
%
Over 97\% of stars end their lives as white dwarfs (WDs). The WD
cooling sequence (CS) of a globular cluster (GC) is shaped by age and
star formation history of that cluster, and provides a unique
opportunity to conduct a census of its already evolved massive star
population. WD\,CSs in old stellar populations can also provide
critical information on how the chemical composition of a stellar
remnant influences its thermal evolution.

\begin{figure*}
\begin{center}
\includegraphics[height=61mm]{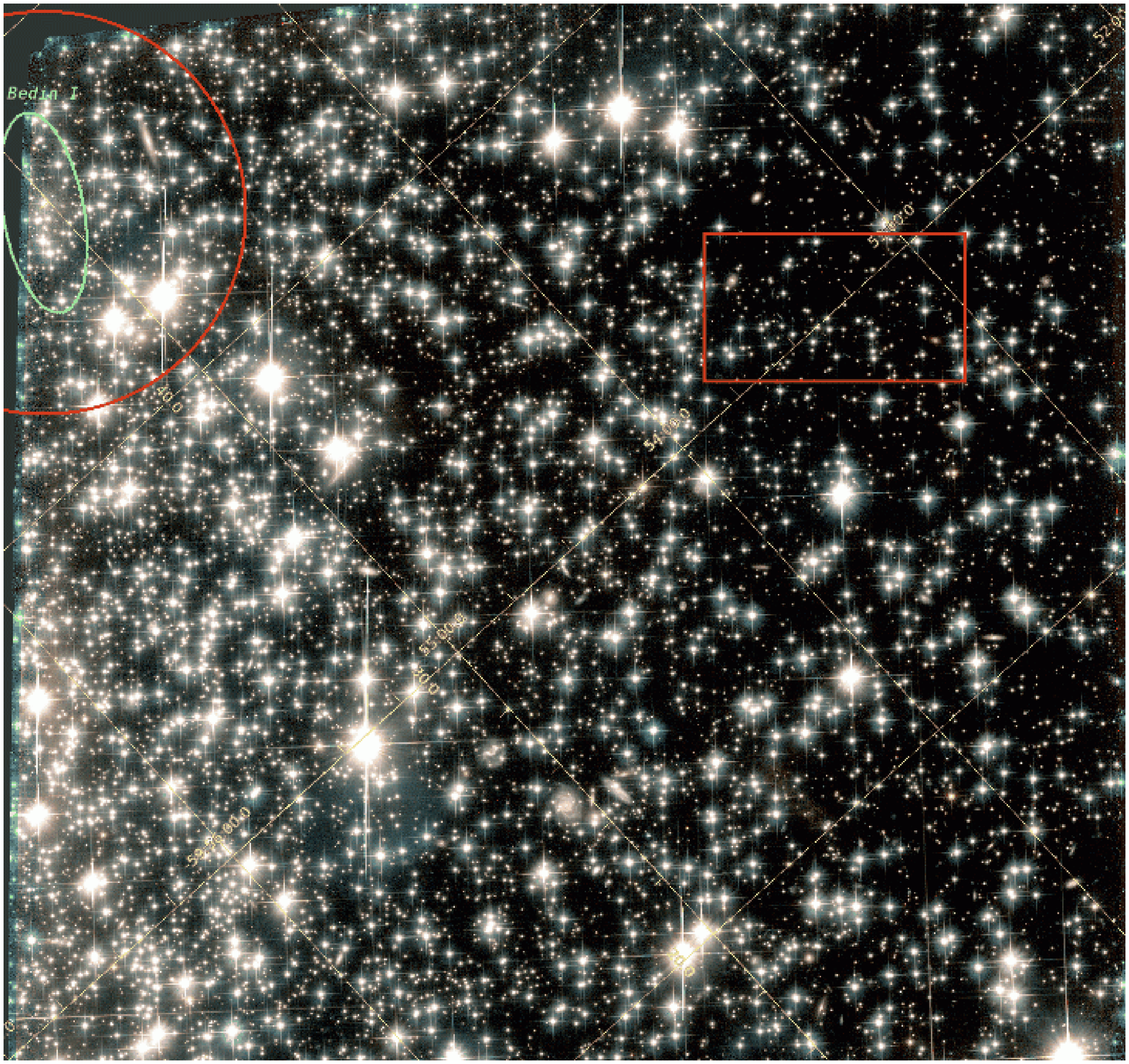}
\includegraphics[height=61mm]{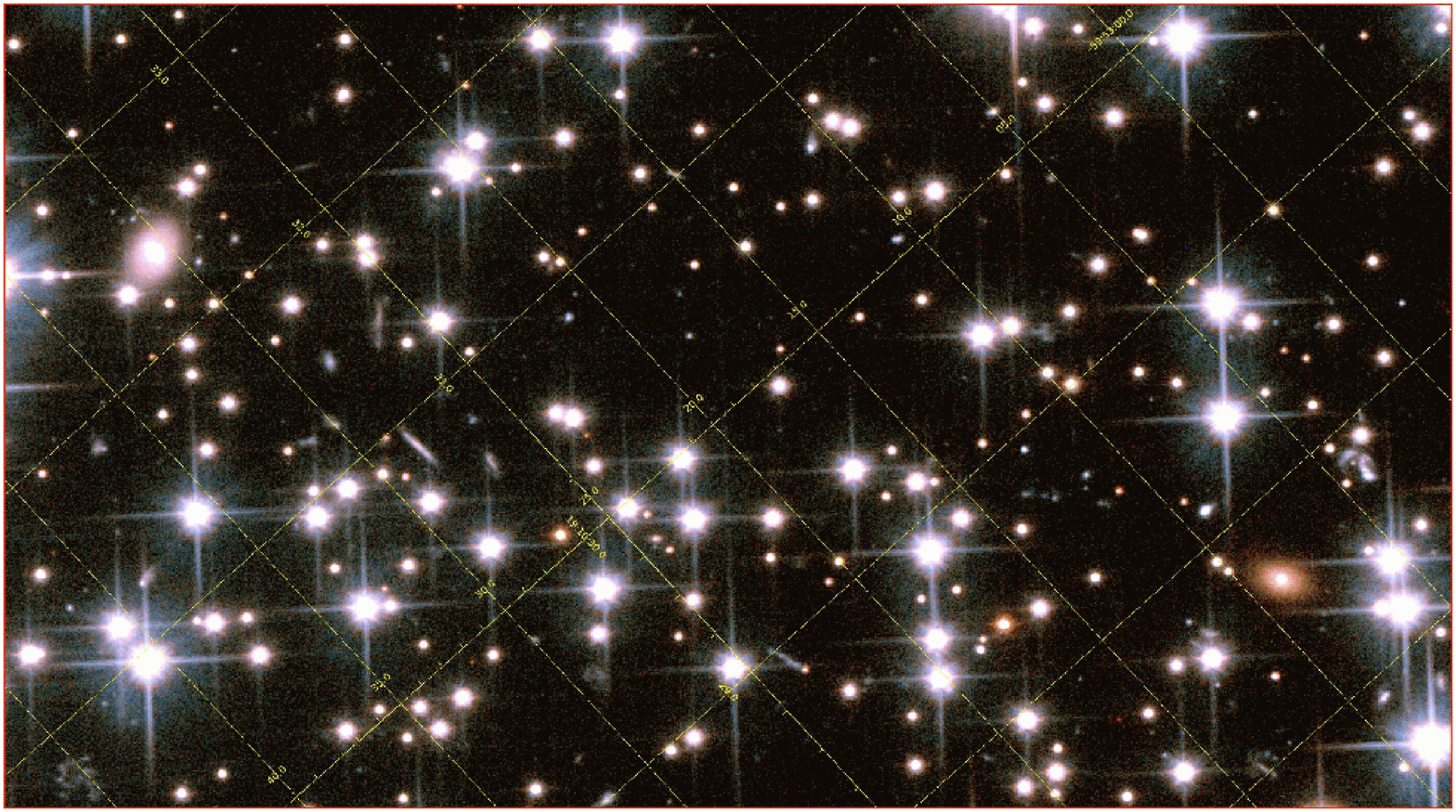}
\caption{
\textit{(Left:)} Trichromatic view of the entire ACS/WFC field of the
\textit{HST} program GO-15096. Note, the relatively high gradient in
stellar density.  Yellow grids give coordinates for the ICRS.  The
green ellipse indicates the \textit{Bedin\,I} dwarf galaxy, and the
red circle the region masked out as potentially contaminated by the
dwarf galaxy.
\textit{(Right:)} Zoom-in of the area indicated with a red rectangular
in the left-panel.
This being representative of the regions with the lowest sky
brightness in the field.
\label{stack}
}
\end{center}
\end{figure*}
%

For the oldest stellar populations, the WD\,CS lies in the faintest
and largely unexplored regions of the colour-magnitude diagram
(CMD). Deep imaging with the \textit{Hubble Space Telescope (HST)} has
for the first time reached the peak of the luminosity distribution of
the WD\,CS in three \textit{classical} GCs, namely: NGC\,6397, M\,4
and 47\,Tucanae [Anderson et al.\,2008a, Bedin et al.\,2009, Kalirai
  et al.\,2012], and has revealed an unexpectedly complex, and
double-peaked, WD\,CS in the metal rich old open cluster NGC\,6791
(Bedin et al.\,2005a, 2008a,b).
Each of the studied GCs hosts multiple stellar populations (mPOPs) and
they show small mean spreads of the initial He abundances (Milone et
al.\ 2018). Their WD\,CSs are consistent with predictions for
single-population systems (Richer et al.\ 2013, Campos et al.\ 2016).

One more \textit{HST} investigation of WD\,CSs is in progress on the
massive GC $\omega$\,Centauri ($\omega$\,Cen), where at least 15
sub-populations are known to exist (Bellini et al.\ 2017).  The
massive $\omega$\,Cen has also long been known to host mPOPs, with a
large spread in both [Fe/H] and (important for that investigation)
initial Helium abundance, as deduced from its divided main sequence
(MS, Bedin et al.\,2004, King et al.\,2012).
The upper part of the WD\,CS in $\omega$\,Cen bifurcates into two
sequences (Bellini et al.\ 2013): a blue CS consisting of standard
CO-core WDs, and a red CS consisting of low-mass WDs with both CO and
He-cores. The current hypothesis is that the blue WD\,CS is populated
by the end products of the He-normal stellar population of
$\omega$\,Cen, while the red WD\,CS is populated by the end products
of the He-rich population. Observing the WD\,CS down to the peak of
its luminosity distribution provides a critical test of this
hypothesis, and a resolution of the origin of the multiple WD\,CSs
observed.
That object can help to answer to key questions about He-dependence in
the evolution of WDs.
Fortunately, $\omega$\,Cen is close enough that its entire WD CS is
within the reach of \textit{HST}, and it is the subject of an
investigation in progress (GO-14118+14662 PI: Bedin; Milone et
al.\ 2017, Bellini et al.\ 2018, Libralato et al.\ 2018).

While almost every globular cluster is known to host multiple
populations, every single cluster is unique. The GC NGC\,6752
represents a transition between the relatively simple globular
clusters, and $\omega$\,Cen, the most complex globular cluster
known. NGC\,6752 has an extended blue horizontal branch, a collapsed
core and 3 chemically distinct populations (Milone et al.\ 2010,
2013). It is one of our final opportunities with \textit{HST} to
increase the diversity in our very limited sample of WD\,CSs, thus far
containing only 3 globular clusters, one old open cluster, and the
complex $\omega$\,Cen system.
While the imminent \textit{James Webb Space Telescope} will also be
able to observe the faintest WDs in closest GCs in the IR, currently,
there is no foreseeable opportunity in the post-\textit{HST} era to
observe WD\,CSs in the homogeneous optical photometric system of
\textit{HST}.

%
\section{Observations} 
%

All images for this study were collected with the \textit{Wide Field Channel}
(WFC) of the \textit{Advanced Camera for Surveys} (ACS) of the
\textit{HST} in program GO-15096 (PI: Bedin).
Unfortunately, five out of the planned 40 orbits failed because of
poor guide star acquisition and will be re-observed at a later time
(currently scheduled for August 2019).
Usable data were collected between September 7 and 18, 2018, and
consist of deep exposures of 1270\,s each, 19 in the F814W filter, and
56 in F606W. At the beginning of each orbit shallow images, of
$\sim$45\,s each, were also collected, in total 27 in F606W and only
10 in F814W.
Note that this is an astrometric multi-cycle programme, and a second
epoch (also of 40 orbits) has already been approved (GO-15491) and is
currently scheduled for late 2020.  Proper motions will eventually
provide a near perfect decontamination of NGC\,6752 members from field
objects which are both in the foreground and background of the
cluster.

Paper\,I of this series (Bedin et al.\ 2019) presented the discovery
of a serendipitous dwarf galaxy in the main studied field, while
Paper\,II (Milone et al.\,2019) focused on multiple stellar
populations detected on the low main sequence of NGC\,6752 in our
parallel observations with the \textit{Infra-Red channel} (IR) of the
\textit{Wide Field Camera 3} (WFC3).

%
\section{Data Reduction and Analysis} 
%

All images were pre-processed with the pixel-based correction for
imperfections in the charge transfer efficiency (CTE) with methods
similar to those described in Anderson \& Bedin (2010).
Photometry and relative positions were obtained with the software
tools described by Anderson et al.\ (2008b). In addition to solving
for positions and fluxes, important diagnostic parameters were also
computed, such as the image-shape parameter (\texttt{RADXS},
introduced in Bedin et al.\ 2008a), which quantifies the fraction of
light that a source has outside the predicted point-spread function
(PSF), the local sky noise (\texttt{rmsSKY}, Bedin et al.\,2009),
and the level of crowding (\texttt{o}, Anderson et
al.\ 2008b).
The \texttt{RADXS} parameter is useful for eliminating most of the
faint unresolved galaxies that tend to plague studies of faint point
sources, while the ``\texttt{o}'' parameter tells how much of the flux
within the PSF fitting radius comes from detected and modeled
neighbors with respect to the target source. The parameter
\texttt{rmsSKY} has a minor role in selection of sources, however it
contains precious information on how suitable the surroundings of each
source are for accurate recovery and photometry.

The astrometry was registered to International Celestial Reference
System (ICRS) using sources in common with \textit{Gaia}\,DR2 (Gaia
collaboration 2018) with tabulated proper motions transformed to the
epoch 2018.689826 of \textit{HST} data, following the procedures in
Bedin \& Fontanive (2018).

The photometry from shallow exposures was linked to that from deep
images.  The photometry was calibrated on the ACS/WFC Vega-mag system
following the procedures given in Bedin et al.\,(2005b) using
encircled energy and zero points available at
STScI.\footnote{\texttt{http://www.stsci.edu/hst/acs/analysis/zeropoints}}
Finally, we applied shifts of the order of $\sim$0.02\,mag to link our
photometry to the \textit{state of the art} \textit{HST} photometric
catalogue by Nardiello et al.\ (2018) in F606W and F814W, which is
obtained for a different field centred on the core of the cluster
(Sarajedini et al.\ 2007).  This registration, is important, as it
enables us to have in the same system a better populated upper part of the CMD.
For these calibrated magnitudes we will use the symbols $m_{\rm  F606W}$ and $m_{\rm F814W}$.

Artificial star tests (ASTs) have a fundamental role in this
investigation, as they are used to estimate the completeness level,
the errors, to optimize selection criteria, and to detect the presence
of systematic errors.
We performed ASTs by adding artificial stars to the individual images
as described in Anderson et al.\ (2008b), with a random flat spatial
distribution, a flat luminosity distribution in $m_{\rm F814W}$ between
magnitude 24 and 30, and with colours chosen to place them along the
observed WD CS (a fiducial line defined by hand).
Indeed, following the prescriptions in Bedin et al.\ (2009, Sect.\,
2.3), we used ASTs to perform an \textit{input-output correction},
i.e., to correct the difference between the inserted and recovered
values of magnitudes of artificial stars, a well known systematic
error (often referred to as \textit{stellar migration}) that tends to
increasingly overestimate the fluxes of stars towards fainter
magnitudes.  Basically, stars landing on positive peaks of noise are
preferentially detected but also recovered systematically brighter (as
much as $\sim$0.2\,mag).
Hereafter, our magnitudes for both real and artificial stars are
intended as corrected for such effects.\\

The software presented in Anderson et al.\ (2008b) also corrects for
distortion in all the images and transforms their coordinates to a
common reference frame after removal of cosmic rays and most of the
artifacts.
It then combines them to produce stacked images that, at any location,
give the sigma-clipped mean of the individual values of pixels of all
images at that location.
\begin{figure*}
\begin{center}
\includegraphics[width=168mm]{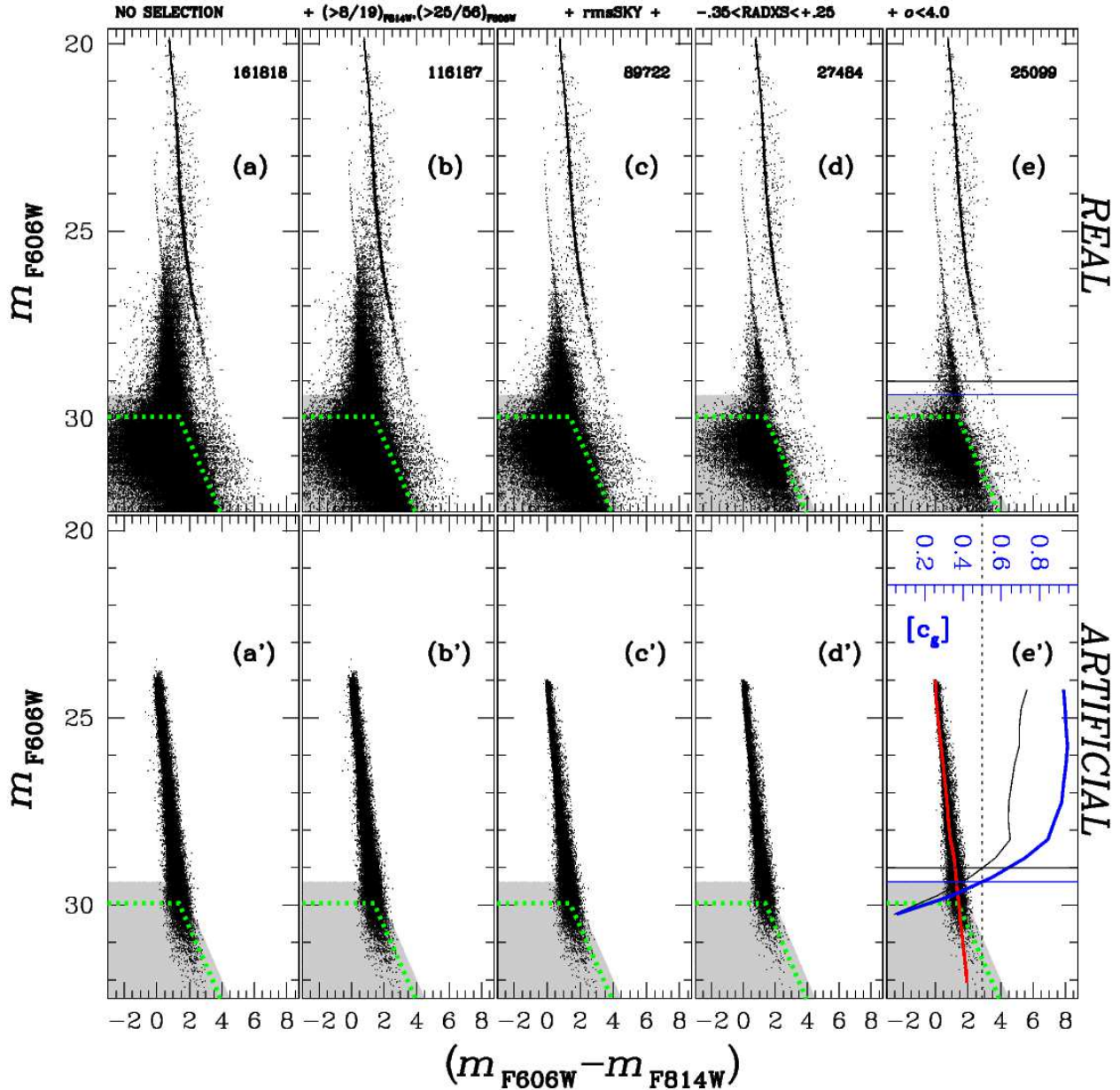}
\caption{
  From left to right, the CMDs are subject to an increasing number of
  selection criteria (see text).  Top panels refer to \textit{real}
  stars while bottom panels to \textit{artificial stars}.  Panel
  (e$^\prime$) also shows the traditional completeness (black line)
  and the effective completeness ([c$_{\rm g}$] in blue, see text),
  the reference scale for which is shown in blue.
\label{CMDs}
}
\end{center}
\end{figure*}
%
In Fig.\,\ref{stack} we show a pseudo-trichromatic stack of the
ACS/WFC field of view (FoV).  [Note that we adopted the F814W images
  for the red channel, the F606W images for the blue channel, and
  computed a wavelength weighted (blue/red$\sim$3) average of the two
  for the green channel.]
As part of this work, we made this image publicly available (with WCS
header linked to \textit{Gaia}\,DR2) as supplementary on-line
electronic material.

%
\section{The Colour-Magnitude Diagram} 
%
%
In this section we demonstrate that we have confidently detected a
sharp peak in the observed luminosity function (LF) of the WD\,CS,
which we will show in the next sections to be consistent with the
peak of the CO WD luminosity distribution in NGC\,6752.

In order to detect the faintest sources we allow 
our algorithms to find any local maximum with a non-null flux in both
filters above the local sky, as a potential real object. In principle,
one every 9 pixels could generate such a peak, we ended up with 'just'
$\sim$162\,000 local maxima, or 1 every $\sim$10$\times$10\,pixel$^2$.
The CMD for these detected local peaks --without any selection-- is
shown in panel (a) of Fig\,\ref{CMDs}.

Paper\,I presented the discovery of a dwarf spheroidal galaxy (dSG)
--designated \textit{Bedin\,I}-- in the background of NGC\,6752, which
is close to the upper-left corner of the FoV in Fig\,\ref{stack}. The
RGB stars of this stellar system happen to contaminate the region of
the CMD where WDs are located.
For this reason we have chosen to completely mask out and not use any
sources in the region of the FoV occupied by this resolved background
dwarf galaxy. The mask is circular, centred on Bedin\,I and has a
radius of 800\,pixels ($\sim$40\,arcsec).
We also require that any of the detected sources generates a peak in
at least 9 out of the 19 F814W images and in at least 26 out of the 56
F606W images.
With these first selections the number of suitable peaks drops to
$\sim$116\,000, and their CMD is shown in panel (b). Still many of
these peaks are artifacts and spurious detections.

Our next selection is illustrated in panel (c) and removes objects
falling in the regions of the field that are too noisy to detect such
objects at all.
As we will see, this mild selection in \texttt{rmsSKY} has more
significant implications when used to estimate the completeness
limited to portions of the FoV suitable for actual detection of faint
sources.

The most effective parameter for the selection of real point sources
is \texttt{RADXS} (Bedin et al.\ 2008a, 2009, 2010, 2015).  This
parameter is a measure of how much flux there is in the pixels just
outside of the core, in excess of the prediction from the PSF:
\texttt{RADXS} is positive if the object is broader than the PSF, and
negative if it is sharper. Even mild selection criteria imposed on
this parameter is sufficient to reject 2/3 of the initially detected
peaks.
The result of this selection of peaks with
$-0.35<\texttt{RADXS}<+0.25$ is shown in panel\,(d).

Our last selection criterion is applied through the parameter
\texttt{o} (Anderson et al.\ 2008b), the fraction of light due to
neighbours.  We require that the flux of light within the PSF
normalized-aperture due to neighbours does not exceed 4 times the flux
of the target source (\texttt{o}$<$4).
This criterion further reduces the sample by $\sim$10\%, and the
$\sim$25\,000 surviving sources are shown in panel (e).\\
 
It is clear, however, that there is only so much we can do to optimize
selection criteria parameters at the faintest end, as going fainter,
it becomes increasingly difficult to estimate the real shape of
sources, or to get any meaningful intrinsic parameters at all.
\sout{This according to the general principle that }\textit{\sout{``all cats are grey in the dark''.}}\footnote{
\sout{The phrase is attributed to Benjamin Franklin.} ;) }
At this point it becomes more important to merely assess whether, and
at what level, a given local peak is significant with respect to the
noise of the local sky.
Therefore we proceed as follows: Given that in filter F814W our PSFs
contains $\sim$16\% of the light in its central pixel, and about
$\sim$18\% in F606W, we can use the standard deviation ($\sigma$) of
pixel values in empty regions of the sky to estimate the magnitude
level that corresponds to 3-$\sigma$ and 5-$\sigma$ levels in the sky
noise.  To transform these noise levels into total\,flux, we divide
these 3-5\,$\sigma$ by the fraction of total flux (i.e., 0.16 for
F814W and 0.18 for F606W), then convert these values into instrumental
magnitudes $-2.5\log{\rm (total\,flux)}$, and finally calibrate them
with ACS/WFC Vega-mag zero points.
As expected, the bulk of these peaks have white noise, and as such an
instrumental colour centred at instrumental magnitudes 0.00 (or once
calibrated, $\sim$1 in $(m_{\rm F606W}-m_{\rm F814W}$).

The grey shaded areas in all panels of Fig.\,\ref{CMDs} show the
regions for peaks with a significance less than 5-$\sigma$, as
estimated in the least noisy portion of the field (the rectangular
region in red in Fig.\,\ref{stack}). The green dotted lines instead,
mark the 3-$\sigma$ significance level.
This essentially is the estimated noise floor-level in the darkest
portion of the FoV.
Hereafter, we choose to consider all the detections below these green
lines as not significant.\\

Panels (a$^\prime$), (b$^\prime$), (c$^\prime$), (d$^\prime$), and
(e$^\prime$) repeat these selections for artificial stars.
However, there are two additional conditions for ASTs: We know where
we added artificial stars, and at which magnitudes. Therefore we
impose generous requirements for the recovered artificial stars. An
artificial star is successfully recovered if it lies within 1 pixel
from its positions in both coordinates, and within 0.75 magnitude from
its added flux in both filters.
%
\begin{figure*}
\begin{center}
\includegraphics[width=88mm]{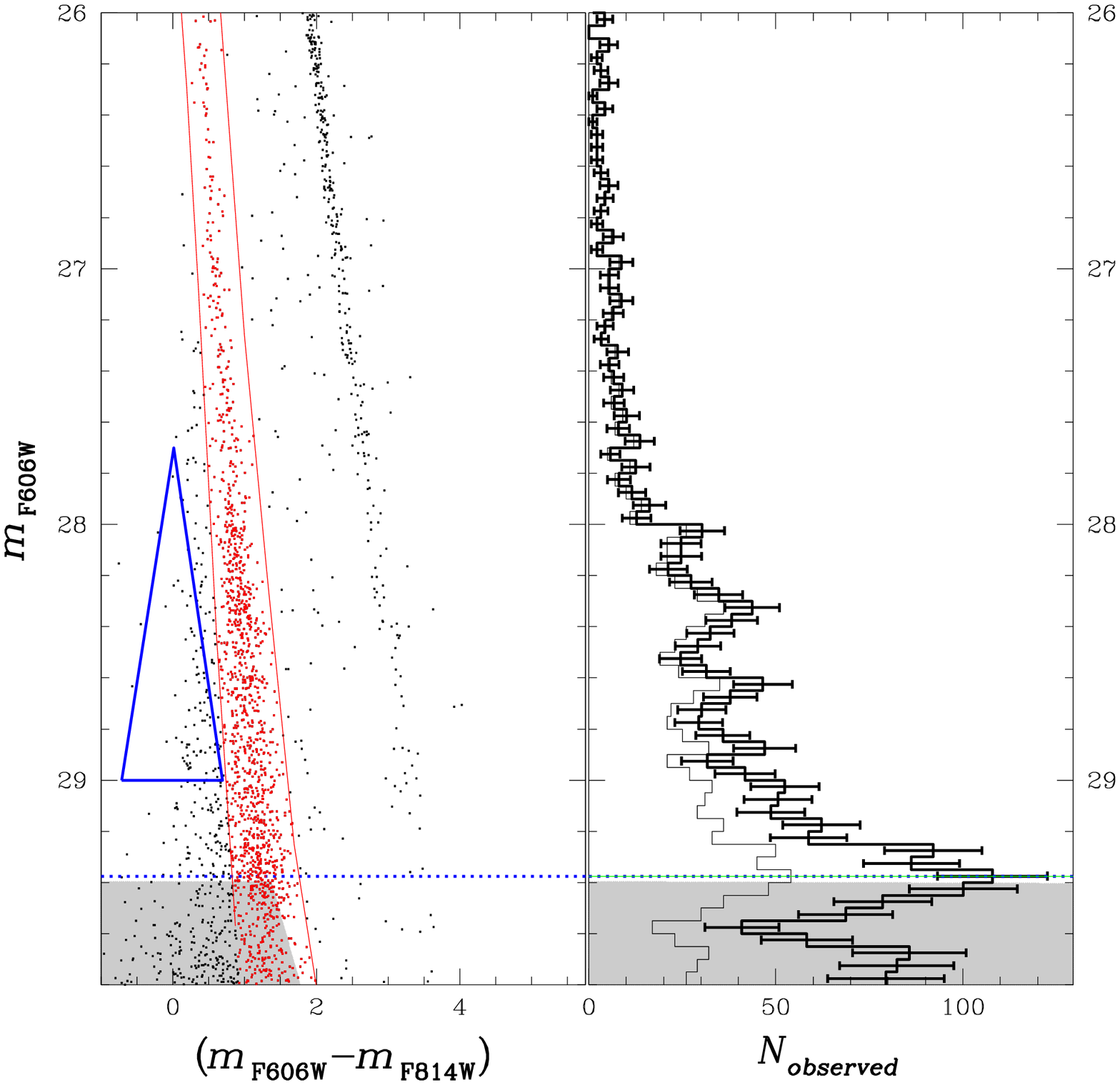}
\includegraphics[width=88mm]{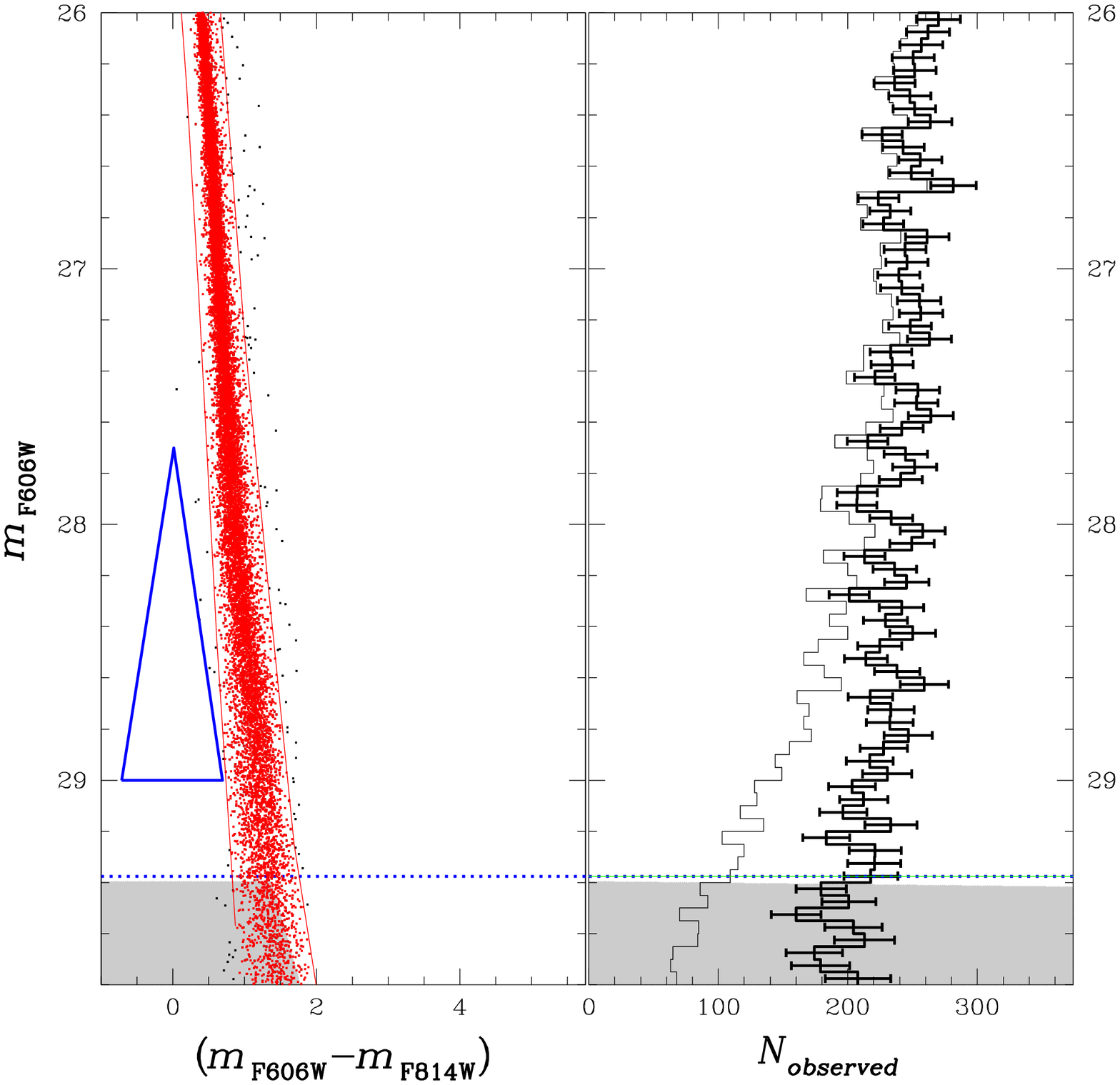}
\caption{
  \textit{(Left Figure:)} on the left panel the CMDs for selected real
  sources.  The blue triangle indicates the region where faint blue
  unresolved background galaxies are expected to be; it was defined by
  Bedin et al.\ (2008a) using data from the \textit{Hubble Ultra Deep Field}.
  The two red lines delimit the region of the CMDs where we assume
  most of WDs to be.
  The right panel show the observed LF for the WD candidates
  highlighted in red in the left panel (thin histogram).  The
  completeness-corrected WD CS LF is shown with error bars (thick
  line).  A horizontal blue dotted-line marks the 50\% of [c$_{\rm g}$],
  while the green horizontal line marks the location of the
  peak of the WD CS LF at $m_{\rm F606W}=29.375\pm0.100$.  The grey
  regions indicate the 5-$\sigma$ threshold above the local sky,
  estimated for the darkest portion of the FoV.
  \textit{(Right Figure:)} same panels just for the artificial stars.
  Note that the two red line encompass most of the recovered stars.
  The artificial stars were added along the WD CS with a flat
  distribution in F814W, which however results in a not flat
  distribution in F606W.
\label{WDCSLF}
}
\end{center}
\end{figure*}
%
%
These criteria are sufficient to remove all the potential mismatches
of sources with local maxima in the noise, and that is the reason why
the bulk of the floor-level noise is not seen in CMDs of ASTs.
These CMDs show that added stars are successfully recovered down to
the estimated 3\,$\sigma$ level.

The concept of effective (or local) completeness was first introduced
in Bedin et al.\ (2008a), and successively used in other independent
studies (Bedin et al.\ 2009, 2010, 2015).
The idea is simple: we cannot hope to recover faint stars around the
bright halos of saturated stars, where the shot-noise is already
higher than the signal from the faint sources we would hope to
recover/detect. Therefore we should limit our search for faint sources
to only those regions of the FoV where the sky noise is sufficiently
faint to allow the faint sources to be detected.

We use ASTs to map these regions and, in practice, we limit our search
for faint stars to only those regions. Naturally, if the completeness
is limited to only suitable search areas that exclude the bright halos
of saturated stars (see Fig.\,\ref{stack}), this approach improves
with respect to the completeness computed for the overall field.
Panel (e$^\prime$) also shows both the traditional (overall)
completeness and the effective (local) completeness, where the
top-axis (in blue) refers to completeness values.  The traditional
completeness is indicated with a black line, and this seems to never
exceed $\sim$70\% for the faint magnitude range considered here.
Instead, the effective completeness (blue line) can be as high as
95\%. Note that the ratio of the two tells us what is the usable
fraction of the area to search for stars of a given magnitude.

The conventional rule of trusting only completeness down to 50\% would
sets the limit of validity of our WD CS LF study down to $m_{\rm F606W}\simeq29.4$.
Interestingly enough, however, this is brighter than the estimated
5-$\sigma$ regions highlighted in grey in Fig.\,\ref{CMDs}.

%
\subsection{The Observed WD\,CS\,LF of NGC\,6752}
%
%
Lacking proper motions until the acquisition of our second epoch, our
next best option is to obtain as pure as possible a sample of
NGC\,6752 WD members based on their position in the CMD.

We do expect several field objects to fall along the WD\,CS locus of
NGC\,6752, and to affect the exact shape of the WD LF we derive here.
However, we cannot foresee a scenario in which the field objects would
introduce a well-defined peak mimicking that of the WD\,LF. In the
following we will review three possible sources of contaminants
(background/foreground field stars; resolved galaxies, and unresolved
galaxies) and address why these sources are very unlikely to introduce
the observed feature.\\

We tested the best available models of star distribution for Galactic
\textit{field stars} in the background and foreground of NGC\,6752
(such as the Besan\c{c}on
models\footnote{https://model.obs-besancon.fr/modele\_descrip.php} by
Robin et al. 2003) and obtained a rather flat distribution with no
peaks. Only $\sim$200 field stars are expected within the WD\,CS
region between $m_{F606W}= 22$ and 29.6, compared to a total of over
1200 observed, with between 5 and 9 stars per 0.05-magnitude bin in
the region of the WD LF peak.

Empirically, these numbers and color-magnitude distributions of field
stars are also supported by similar studies on WDs of star clusters at
any height above the Galactic plane (Bedin et al.\ 2008, 2009, 2010,
2015, Anderson et al.\ 2008a, Kalirai et al.\ 2012). \\

Similarly, we do not expect contamination by background galaxies to
mimick the observed feature. All background galaxies sufficiently
large and bright to have well established shapes (i.e., above the
5$\sigma$ lines) were easily removed by the selection in the
\texttt{RADXS} parameter (even sharp quasar-like object, at
\textit{HST} resolution, reveal departures from PSFs, e.g., Bedin et
al.\ 2003).\\


A possible concern could be the contamination by faint, compact, blue
and unresolved-galaxies in the background of NGC\,6752.\footnote{Those
  not already removed by the \texttt{RADXS} photometric selections
  described in previous section.}
However, thanks to decades of deep \textit{HST} observations their
numbers and loci in the CMD are well known.  For this purpose, a
region in the CMD was carefully defined by Bedin et al.\,(2008a, 2009)
and demonstrated to not overlap with the clusters WD\,CSs.

In Fig.\,\ref{WDCSLF} we show the CMDs for the selected sources
defined in panels (e) and (e$^\prime$) of Fig.\,\ref{CMDs}.
Again, ASTs turn out to be very useful to define the location of
reliably measured WDs along the fiducial CS of NGC\,6752 for the
entire magnitude range of interest.  Sources between the two red lines
defined by-hand in the CMD for ASTs (right panels) are assumed to be
good WD members of NGC\,6752 including photometric errors.
Identical lines are also used to define the sample of WDs among real
stars (left panels).
The blue triangle shows the location of blue-compact unresolved background
galaxies adopted by Bedin et al.\ (2009) using \textit{Hubble Ultra
  Deep Field} (HUDF) data, and transformed into the F814W filter as was done
in Bedin et al.\ (2008a). Note that the adopted WD-region is not
contaminated by this blue-compact galaxies triangular region; this is
a conservative statement, as the lower reddening of NGC\,6752 with
respect to NGC\,6791 would make this triangular region about 0.1\,mag
bluer (and so even more distant from the WD-region).
Note also, that these selections are large enough to take into account
the photometric error distributions of unresolved point sources.

On the CMD for real sources of Fig.\,3, there are many sources that
lie outside the red boundary that defines the white dwarf sequence.
These are the expected mixture of stars in the Galactic field, many of
which are WDs.  These field stars are expected to contaminate both the
triangular region of the blue-compact galaxies and the WD region,
therefore, they affect the exact shape of the WD\,LF.

We then count the observed objects ($N$) at the various $m_{\rm F606W}$
magnitudes, and correct their values for the \textit{effective}-completeness
($c_g$), obtaining the completeness-corrected WD\,LF values ($N_{c_g}$).
In Table\,1 we report the values of the entire WD LF and relative
uncertainties.
We must treat these numbers with care, however.  Not only because
energy equipartition is expected to cause the more massive WDs to
migrate toward the centre of the cluster, but also because of
unaccounted for residual contamination due to field objects or to
artifacts.
Nevertheless, these uncertainties can only bias the relative numbers,
while the WD\,LF peaks at a magnitude where our completeness is still
reliable, and where contaminants cannot affect its true positions.

We define the peak of the WD\,LF as the bin with the highest value,
which is at $m_{\rm F606W}$$\simeq$29.4. As formal uncertainty we
assume $\sigma$=FWHM/2.354$\simeq$0.1\,mag, where FWHM is the
full-width half maximum of the peak, which is 0.2\,mag.
At the magnitude of the peak, we observe a rather flat distribution of
field objects on the blue-side of the WD\,CS region, at maximum
$\sim$20\,objects/bin (completeness corrected).  Assuming a similar
field contamination with a similar distribution within the WD\,CS
region (the best we can do), we do not expect the position of the peak
to change.
%

%
\section{Theoretical assessment of the LF and conclusions} 
%
%
We now present the first preliminary comparison of the WD LF with models.

\begin{figure*}
\includegraphics[width=18truecm]{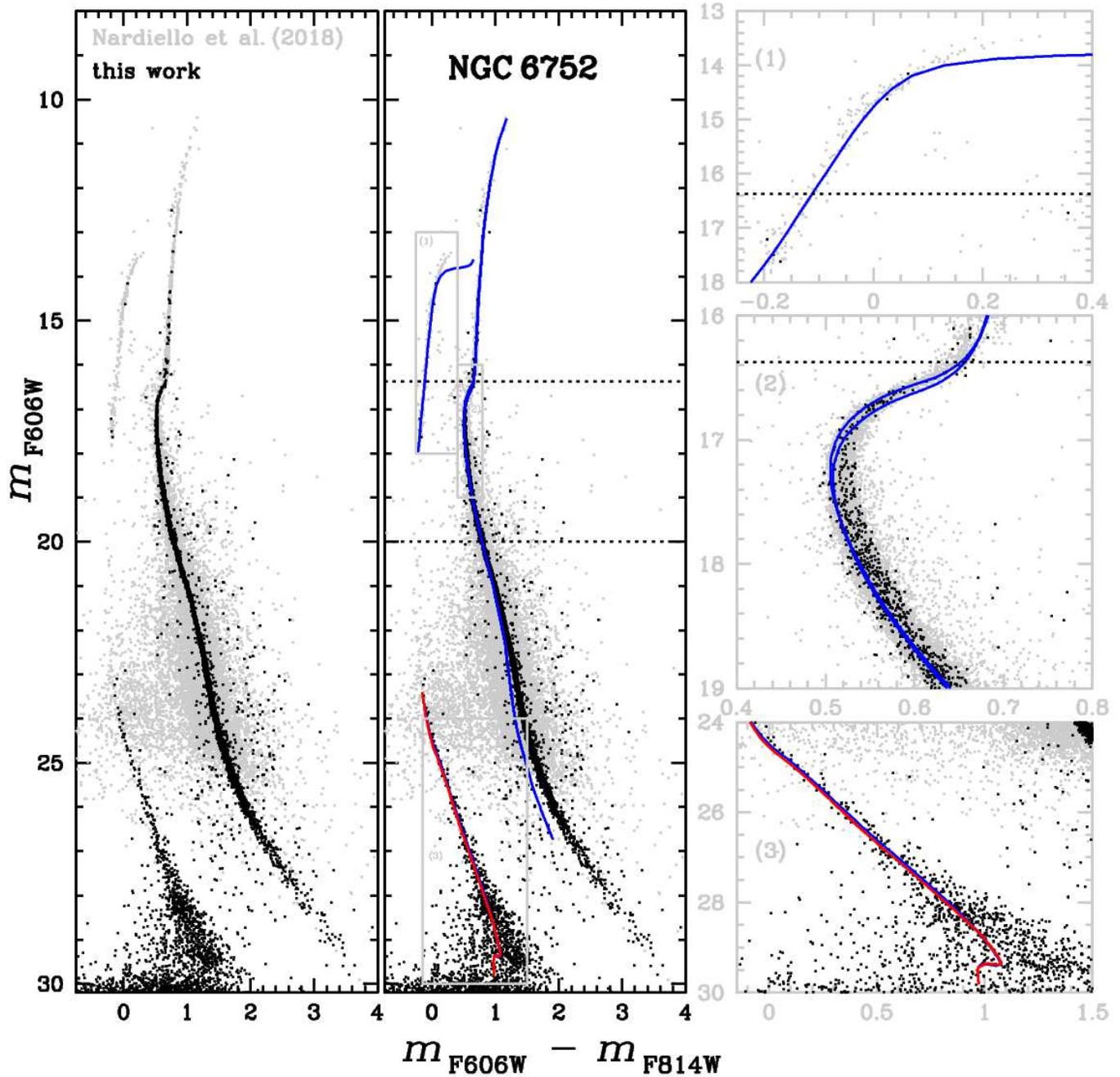}
\caption{
  \textit{(Left:) }
  Entire CMD of NGC\,6752, including main sequence, subgiant, red
  giant, horizontal branches and WDs.  Black dots denote our
  photometry, while grey dots the Nardiello et al.\ (2018).
  \textit{(Middle:) }
  the same CMD, where solid lines display: isochrones for 13 and
  14~Gyr for the MS+TO, two 14~Gyr DA WD isochrones and the
  theoretical zero age horizontal branch, all shifted by $E(B-V)$=0.05
  and $(m-M)_0$=13.10 (see text for details).
  In this panel the horizontal dotted lines mark where saturation start
  in our deep and short exposures. 
  \textit{(Right:) }
  Zoom-in of the three grey regions defined in the middle panel (and
  labeled as (1), (2) and (3)) to highlight the comparison of  
  observations with theory in the different evolutionary phases.
}
\label{MSfit}
\end{figure*}
%

Future epochs will enable us to carry out an exhaustive
characterization of the observed WD LF of NGC\,6752 and of its exact
shape, thanks to improved S/N (through the doubling of the exposure
time) and proper motion cleaning for artifacts and for field objects
(in both the background and foreground).
Indeed, the increased S/N planned for the future, will enable us to
study in detail not only the exact shape of the WD\,LF but
\textit{also} the exact shape of the WD\,CS in the CMD (Anderson et
al.\ 2008a).
Here, \textit{instead}, we only aim to just verify that the rather
sharp peak in the observed LF, as supported by the artificial star
tests, is roughly consistent with the magnitude theoretically
expected for the peak in the LF of the cluster WD\,CS.\\

As a first step we have determined the reddening and cluster distance
modulus from our data, by fitting theoretical isochrones to the colour
of the cluster unevolved main sequence between $m_{\rm F606W}$=18 and
21, and the red giant branch (see middle panel of Fig.~\ref{MSfit}).
To this purpose we have employed our own photometry, complemented by
the Nardiello et al.\ (2018) catalogue, which is consistent with our
photometry for the common evolutionary phases. This latter catalogue
includes a well populated red giant branch, as well as the horizontal
branch.  Stars in this cluster display a small mean range of initial
helium mass fraction ($\Delta Y$=0.015$\pm$0.005, Milone et al.\ 2018;
Nardiello et al.\ 2015; Paper\,II), so that we can proceed by
employing isochrones with a single initial He abundance without
introducing any major bias in our analysis.

%
\begin{figure}
\includegraphics[width=\columnwidth]{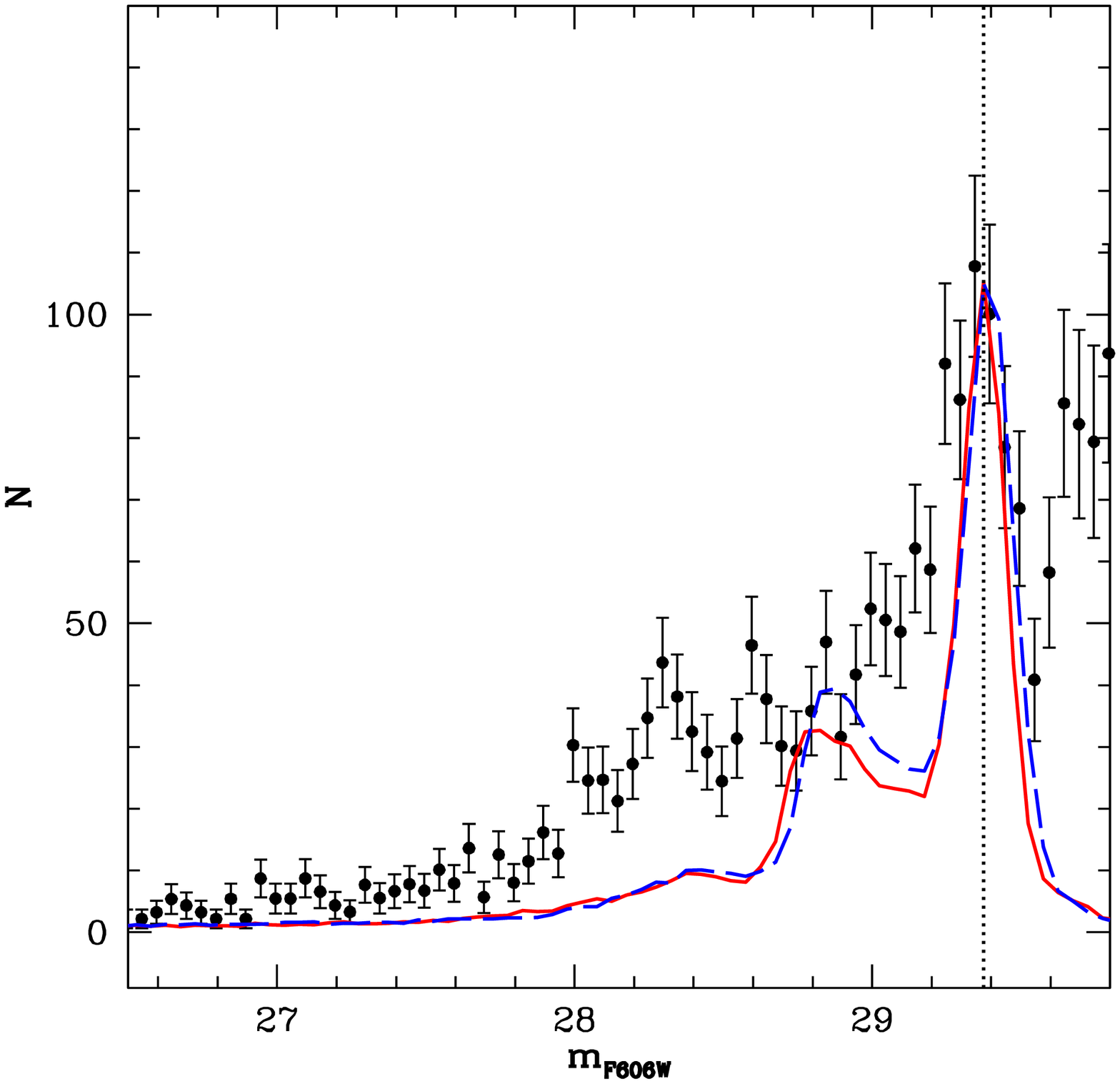}
\caption{ Observed (completeness corrected) WD LF (filled circles with
  error bars), together with two different theoretical LFs (red solid
  and blue dashed lines) for an age of 14~Gyr, shifted by our derived
  cluster distance modulus and extinction. The solid line corresponds
  to the LF calculated with isochrone~A of Fig.~\ref{MSfit}, whilst
  the dashed line corresponds to calculations employing isochrone~B in
  the same figure.  The dotted vertical line marks the magnitude level
  of the 50\% completeness threshold.  }
\label{lffit}
\end{figure}
%

We employed the $\alpha$-enhanced BaSTI (Pietrinferni et al.\ 2006)
isochrones with [Fe/H]=$-$1.6, $Y$=0.246, and employed the extinction
law in the ACS filters determined by Bedin et al.\,(2005b). The
isochrone metallicity is close to [Fe/H]=$-$1.48$\pm$0.06 which was
determined spectroscopically by Gratton et al.\ (2005) for NGC\,6752.
The fit provides $E(B-V)$=0.05 (consistent with
$E(B-V)$=0.046$\pm$0.005 as determined by Gratton et al.\ 2005) and
$(m-M)_0$=13.10. This distance modulus compares well with $(m-M)_{0}
=$12.92$\pm$0.24 obtained from the \textit{Gaia}\,DR2 cluster parallax
suggested by the Gaia collaboration (2018) after the offset correction
discussed by Lindegren et al.\ (2018).  The error on this
\textit{Gaia} distance modulus is, however, still sizable, due to the
\lq{calibration noise}\rq\ described by the Gaia collaboration (2018).
The top section (1) of the right panel of Fig.\,\ref{MSfit} shows that,
with this distance modulus and reddening, the theoretical zero age
horizontal branch sequence is also a good match to the observed
counterpart (the redder part of the observed HB is populated by stars
evolving to the asymptotic giant branch phase).  Ages of 13 and 14~Gyr
bracket the position of the age-sensitive subgiant branch in the CMD
(see the middle section of the same panel).
The isochrones do not include the effect of atomic diffusion.  Its
inclusion would decrease these ages by $\sim 1$~Gyr (see, e.g.,
Vandenberg et al. 2002).

The bottom section (3) of the left panel of Fig.\,\ref{MSfit} compares
the observed CS with BaSTI WD isochrones (Salaris et al.\,2010),
shifted employing the reddening and distance modulus derived above
from the MS-TO isochrones fit.
These WD isochrones include CO separation upon cristallization by
employing the Segretain \& Chabrier (1993) phase diagram for the CO
mixture. We have determined with appropriate calculations that the
more modern Horowitz et al.\ (2010) CO phase diagram would decrease
the WD cluster ages by $\sim$0.9\,Gyr, compared to the results reported
below.
We show two DA isochrones for an age of 14~Gyr, calculated with
progenitor lifetimes (and CO profiles) for the same chemical
composition as the MS-TO isochrones.
These WD isochrones cover the WD mass range from 0.54 to
1.0\,$M_{\odot}$.  More massive WDs between 1.0\,Mo and
$\sim$1.4\,$M_{\odot}$ (if they form in metal poor clusters) would be
fainter than the faintest magnitude of our isochrones. However, for
any reasonable mass function, they would constitute only a fraction of
a percent of the total cluster WD population.
The first isochrone (displayed in red) has been calculated with the
linear initial-final mass relationship (IFMR) from Salaris et
al.\ (2009).
The IFMR was extrapolated linearly such that initial masses typical of
stars populating Galactic globular cluster main sequence turn off
($\sim$0.8$M_{\odot}$) produce WDs with mass 0.54$M_{\odot}$ (this
IFMR was determined down to values of initial masses
$\sim$1.8$M_{\odot}$).
This mass is consistent with the observational analysis of Kalirai et
al.\, (2009), who show that bright WDs in the Galactic globular
cluster M\,4 are generally of DA type, with a typical mass equal to
0.53$\pm$0.01$M_{\odot}$ (we denote this isochrone as
\textit{isochrone A}).

The second DA isochrone (displayed as blue solid line in the figure)
is calculated by extrapolating Salaris et al.\ (2009) IFMR to reach a
WD mass equal to 0.49$M_{\odot}$ for initial masses below
1.0$M_{\odot}$ (we denote this isochrone as \textit{isochrone B}).  The
reason for this choice is that, as derived from the detailed
horizontal branch modelling by Cassisi et al.\ (2014), 
a fraction of the stars populating this cluster's blue HB, have a
mass close to the core mass at the He-flash for this metallicity
(equal to $\sim$0.49$M_{\odot}$). They will evolve straight to the WD
phase skipping the asymptotic giant branch.

Isochrones A and B differ only at magnitudes brighter than
$m_{F606W}\sim$29.0, where the CS is populated by fast evolving WDs
produced by stars that have just left either the HB, or the
post-asymptotic giant branch phase.  Overall, the observed CS is
closely matched by the theoretical isochrones at magnitudes brighter
than $m_{F606W}\sim$28.2 , which however progressively diverge towards
bluer colours with increasing magnitudes.

Figure~\ref{lffit} compares the empirical (completeness corrected) WD
LF with two theoretical LFs determined from the two 14\,Gyr old DA WD
isochrones of Fig.\,\ref{MSfit}. Random photometric errors were added
to stars, following the results of the artificial star tests, and
shifted by the derived extinction and distance modulus.

The total number of stars in the theoretical LFs have been chosen to
approximately match the value of the peak of the observed LF at
$m_{\rm F606W} \sim$29.4, employing a WD progenitor mass function as a
power law with exponent $-$1.9.  This exponent was selected to match
approximately the observed star counts at least at another magnitude
level ($m_{\rm F606W} \sim$28.8), in addition to the peak at $m_{\rm
  F606W} \sim$29.4. Varying the exponent of the progenitor mass
function alters the shape of the theoretical LF, but not the
brightness of its peak.

The two different IFMRs employed for isochrone~A and B produce
identical LFs at the faint parts of the CS.  This is not surprising
given that the main difference is the value of the initial mass only
for low-mass progenitors.\\~

An exhaustive comparison of observed versus theoretical LFs must wait
for future epochs that will enable to \textit{clean} the cluster WD
sample more reliably.  As things stand, irrespective of the choice of
the progenitor mass function exponent, the theoretical LFs cannot
match the observed LF along the entire CS.  The purpose of this
comparison is therefore just qualitative, but it serves to highlight
the fact that the magnitude level of the predicted peak of the WD LF
does match the magnitude of the observed LF peak, when an age
consistent with the cluster TO age is employed.\\~

This strengthens the case for our detection of the peak in the WD\,LF
of NGC\,6752.

%
\section*{Acknowledgments}
This work is based on observations with the NASA/ESA Hubble Space
Telescope, obtained at the Space Telescope Science Institute, which is
operated by AURA, Inc., under NASA contract NAS 5-26555.
J.A., R.M.R., D.A., Ad.B, An.B, M.L, and acknowledge support from
\textit{HST}-GO-15096.
A.P.M. acknowledges funding from the European Research Council (ERC)
under the European Union's Horizon 2020 research innovation programme
(Grant Agreement ERC-StG 2016, No 716082 `GALFOR'.
A.P.M.\, acknowledges support from MIUR through the FARE project
R164RM93XW ‘SEMPLICE’.
A.F.M.\, has received funding from the European Union’s Horizon 2020
research and innovation programme under the Marie Sk{\l}odowska-Curie
Grant Agreement No.\,[797100].
D.N.\ acknowledges partial support by the University of Padova,
Progetto di Ateneo, Grant BIRD178590.
This work is supported in part by the NSERC Canada (P.B.).
L.R.B.\ acknowledges support by MIUR under PRIN program \#2017Z2HSMF,
...and reminds in this astro-ph version that July 16$^{\rm th}$ is
also L.R.B. B-day!!! ;)

\newpage

\begin{table*}
\caption{
Completeness-corrected  white dwarf  luminosity function (see text).
}
\begin{tabular}[h]{ccccc|ccccc|ccccc} 
\hline\hline
& & & & & & & & & & & & & & \\
$m_{\rm F606W}$ & $N_{c_g}$ & $\sigma_{N_{c_g}}$ & $N$ & [c$_{\rm g}$] & 
$m_{\rm F606W}$ & $N_{c_g}$ & $\sigma_{N_{c_g}}$ & $N$ & [c$_{\rm g}$] & 
$m_{\rm F606W}$ & $N_{c_g}$ & $\sigma_{N_{c_g}}$ & $N$ & [c$_{\rm g}$] \\ 
& & & & & & & & & & & & & & \\
\hline
 24.025 &  1.0863 & 1.0863 & 1 & 0.9205 & 26.025 &  4.2538 & 2.1269 & 4 & 0.9403 &  28.025 & 30.3241 & 5.9470 &26 & 0.8574 \\                                
 24.075 &  0.0000 & 0.0000 & 0 & 0.9212 & 26.075 &  0.0000 & 0.0000 & 0 & 0.9397 &  28.075 & 24.5926 & 5.3665 &21 & 0.8539 \\                                
 24.125 &  1.0846 & 1.0846 & 1 & 0.9220 & 26.125 &  5.3244 & 2.3811 & 5 & 0.9391 &  28.125 & 24.6935 & 5.3886 &21 & 0.8504 \\                                
 24.175 &  1.0838 & 1.0838 & 1 & 0.9227 & 26.175 &  2.1312 & 1.5070 & 2 & 0.9384 &  28.175 & 21.2531 & 5.0094 &18 & 0.8469 \\                                
 24.225 &  1.0829 & 1.0829 & 1 & 0.9234 & 26.225 &  3.1989 & 1.8469 & 3 & 0.9378 &  28.225 & 27.2691 & 5.6860 &23 & 0.8434 \\                                
 24.275 &  1.0821 & 1.0821 & 1 & 0.9242 & 26.275 &  5.3369 & 2.3867 & 5 & 0.9369 &  28.275 & 34.6987 & 6.4434 &29 & 0.8358 \\                                
 24.325 &  2.1624 & 1.5291 & 2 & 0.9249 & 26.325 &  1.0688 & 1.0688 & 1 & 0.9356 &  28.325 & 43.6949 & 7.2825 &36 & 0.8239 \\                                
 24.375 &  0.0000 & 0.0000 & 0 & 0.9256 & 26.375 &  4.2811 & 2.1405 & 4 & 0.9344 &  28.375 & 38.1762 & 6.8566 &31 & 0.8120 \\                                
 24.425 &  2.1590 & 1.5266 & 2 & 0.9264 & 26.425 &  1.0717 & 1.0717 & 1 & 0.9331 &  28.425 & 32.4937 & 6.3725 &26 & 0.8002 \\                                
 24.475 &  0.0000 & 0.0000 & 0 & 0.9271 & 26.475 &  2.1463 & 1.5177 & 2 & 0.9318 &  28.475 & 29.1773 & 6.0839 &23 & 0.7883 \\                                
 24.525 &  1.0778 & 1.0778 & 1 & 0.9278 & 26.525 &  2.1492 & 1.5197 & 2 & 0.9306 &  28.525 & 24.4714 & 5.6141 &19 & 0.7764 \\                                
 24.575 &  0.0000 & 0.0000 & 0 & 0.9285 & 26.575 &  2.1521 & 1.5218 & 2 & 0.9293 &  28.575 & 31.3912 & 6.4077 &24 & 0.7645 \\                                
 24.625 &  1.0761 & 1.0761 & 1 & 0.9293 & 26.625 &  3.2326 & 1.8663 & 3 & 0.9281 &  28.625 & 46.5008 & 7.8601 &35 & 0.7527 \\                                
 24.675 &  1.0753 & 1.0753 & 1 & 0.9300 & 26.675 &  5.3950 & 2.4127 & 5 & 0.9268 &  28.675 & 37.7967 & 7.1429 &28 & 0.7408 \\                                
 24.725 &  3.2233 & 1.8609 & 3 & 0.9307 & 26.725 &  4.3218 & 2.1609 & 4 & 0.9255 &  28.725 & 30.1810 & 6.4346 &22 & 0.7289 \\                                
 24.775 &  0.0000 & 0.0000 & 0 & 0.9314 & 26.775 &  3.2454 & 1.8737 & 3 & 0.9244 &  28.775 & 29.3904 & 6.4135 &21 & 0.7145 \\                                
 24.825 &  1.0729 & 1.0729 & 1 & 0.9320 & 26.825 &  2.1660 & 1.5316 & 2 & 0.9234 &  28.825 & 35.8392 & 7.1678 &25 & 0.6976 \\                                
 24.875 &  0.0000 & 0.0000 & 0 & 0.9327 & 26.875 &  6.5053 & 2.6558 & 6 & 0.9223 &  28.875 & 47.0173 & 8.3116 &32 & 0.6806 \\                                
 24.925 &  0.0000 & 0.0000 & 0 & 0.9333 & 26.925 &  2.1709 & 1.5350 & 2 & 0.9213 &  28.925 & 31.6437 & 6.9052 &21 & 0.6636 \\                                
 24.975 &  1.0707 & 1.0707 & 1 & 0.9339 & 26.975 &  8.6931 & 3.0735 & 8 & 0.9203 &  28.975 & 41.7517 & 8.0351 &27 & 0.6467 \\                                
 25.025 &  3.2100 & 1.8533 & 3 & 0.9346 & 27.025 &  5.4393 & 2.4325 & 5 & 0.9192 &  29.025 & 52.4042 & 9.1224 &33 & 0.6297 \\                                
 25.075 &  0.0000 & 0.0000 & 0 & 0.9352 & 27.075 &  5.4454 & 2.4353 & 5 & 0.9182 &  29.075 & 50.5908 & 9.0864 &31 & 0.6128 \\                                
 25.125 &  2.1372 & 1.5112 & 2 & 0.9358 & 27.125 &  8.7224 & 3.0838 & 8 & 0.9172 &  29.125 & 48.6741 & 9.0385 &29 & 0.5958 \\                                
 25.175 &  1.0679 & 1.0679 & 1 & 0.9365 & 27.175 &  6.5492 & 2.6737 & 6 & 0.9161 &  29.175 & 62.1934 &10.3656 &36 & 0.5788 \\                                
 25.225 &  0.0000 & 0.0000 & 0 & 0.9371 & 27.225 &  4.3710 & 2.1855 & 4 & 0.9151 &  29.225 & 58.7314 &10.2238 &33 & 0.5619 \\                                
 25.275 &  2.1328 & 1.5081 & 2 & 0.9377 & 27.275 &  3.2870 & 1.8977 & 3 & 0.9127 &  29.275 & 92.1133 &13.0268 &50 & 0.5428 \\                                
 25.325 &  0.0000 & 0.0000 & 0 & 0.9384 & 27.325 &  7.7016 & 2.9109 & 7 & 0.9089 &  29.325 & 86.2680 &12.8601 &45 & 0.5216 \\                                
 25.375 &  1.0650 & 1.0650 & 1 & 0.9390 & 27.375 &  5.5243 & 2.4705 & 5 & 0.9051 &  29.375 &107.9029 &14.6837 &54 & 0.5005 \\                                
 25.425 &  1.0642 & 1.0642 & 1 & 0.9396 & 27.425 &  6.6571 & 2.7177 & 6 & 0.9013 & {\it 29.425} & {\it100.1523} & {\it 14.4557} & {\it 48} & {\it  0.4793} \\
 25.475 &  1.0635 & 1.0635 & 1 & 0.9403 & 27.475 &  8.9136 & 3.1515 & 8 & 0.8975 & {\it 29.475} & {\it 78.5872} & {\it 13.0979} & {\it 36} & {\it  0.4581} \\
 25.525 &  1.0628 & 1.0628 & 1 & 0.9409 & 27.525 &  6.7137 & 2.7408 & 6 & 0.8937 & {\it 29.525} & {\it 68.6640} & {\it 12.5363} & {\it 30} & {\it  0.4369} \\
 25.575 &  3.1862 & 1.8396 & 3 & 0.9416 & 27.575 & 10.1135 & 3.3712 & 9 & 0.8899 & {\it 29.575} & {\it 40.8919} & {\it  9.9177} & {\it 17} & {\it  0.4157} \\
 25.625 &  0.0000 & 0.0000 & 0 & 0.9422 & 27.625 &  7.8998 & 2.9858 & 7 & 0.8861 &       &&&&\\
 25.675 &  0.0000 & 0.0000 & 0 & 0.9428 & 27.675 & 13.6008 & 3.9262 &12 & 0.8823 &       &&&&\\
 25.725 &  1.0599 & 1.0599 & 1 & 0.9435 & 27.725 &  5.6915 & 2.5453 & 5 & 0.8785 &       &&&&\\
 25.775 &  5.2995 & 2.3700 & 5 & 0.9435 & 27.775 & 12.5735 & 3.7911 &11 & 0.8749 &       &&&&\\
 25.825 &  1.0606 & 1.0606 & 1 & 0.9429 & 27.825 &  8.0334 & 3.0363 & 7 & 0.8714 &       &&&&\\
 25.875 &  2.1226 & 1.5009 & 2 & 0.9422 & 27.875 & 11.5224 & 3.6437 &10 & 0.8679 &       &&&&\\
 25.925 &  2.1241 & 1.5019 & 2 & 0.9416 & 27.925 & 16.1965 & 4.3287 &14 & 0.8644 &       &&&&\\
 25.975 &  1.0627 & 1.0627 & 1 & 0.9410 & 27.975 & 12.7774 & 3.8525 &11 & 0.8609 &       &&&&\\
%
%
%
\hline 
\hline
\label{WDLFtab}
\end{tabular}
\end{table*}

%


\label{lastpage}



\begin{thebibliography}{}


\bibitem[Anderson, J. et al. 2008a]{2008AJ....135.2114A} Anderson, J. et al. 2008a, AJ, 135, 2114

\bibitem[Anderson et al.(2008b)]{2008AJ....135.2055A} Anderson, J., Sarajedini, A., Bedin, L.~R., King, I.~R., Piotto, G., Reid, I.~N., Siegel, M., Majewski, S.~R., Paust, N.~E.~Q., Aparicio, A., Milone, A.~P., Chaboyer, B., \& Rosenberg, A.\ 2008b, AJ, 135, 2055

\bibitem[Anderson \& Bedin(2010)]{2010PASP..122.1035A} Anderson, J., \& Bedin, L.~R.\ 2010, PASP, 122, 1035

\bibitem[Bedin et al.(2003)]{2003AJ....126..247B} Bedin, L.~R. et al. 2003, AJ, 126, 247

\bibitem[Bedin et al.(2004)]{2004ApJ...605L.125B} Bedin, L.~R., Piotto, G., Anderson, J., et al.\ 2004, ApJL, 605, L125 

\bibitem[Bedin et al.(2005a)]{2005ApJ...624L..45B} Bedin, L.~R. et al. 2005a, ApJL, 624, L45
  
\bibitem[Bedin et al.(2005b)]{2005MNRAS.357.1038B} Bedin, L.~R., Cassisi, S., Castelli, F., Piotto, G., Anderson, J., Salaris, M., Momany, Y., \& Pietrinferni, A.\ 2005b, MNRAS, 357, 1038

\bibitem[Bedin et al.(2008a)]{2008ApJ...678.1279B} Bedin, L.~R., King, I.~R., Anderson, J., et al.\ 2008a, ApJ, 678, 1279 

\bibitem[Bedin et al.(2008b)]{2008ApJ...679L..29B} Bedin, L.~R., Salaris, M., Piotto, G., et al.\ 2008b, ApJL, 679, L29

\bibitem[Bedin et al.(2009)]{2009ApJ...697..965B} Bedin, L.~R., Salaris, M., Piotto, G., et al.\ 2009, ApJ, 697, 965

\bibitem[Bedin et al.(2010)]{2010ApJ...708L..32B} Bedin, L.~R., Salaris, M., King, I.~R., et al.\ 2010, ApJL, 708, L32 

\bibitem[Bedin et al.(2015)]{2015MNRAS.448.1779B} Bedin, L.~R., Salaris, M., Anderson, J., et al.\ 2015, MNRAS, 448, 1779   

\bibitem[Bedin \& Fontanive(2018)]{2018MNRAS.481.5339B} Bedin, L.~R., \& Fontanive, C.\ 2018, MNRAS, 481, 5339

\bibitem[Bedin et al.(2019)]{2019MNRAS.484L..54B} Bedin, L.~R., Salaris, M., Rich, R.~M., et al.\ 2019, MNRAS Letter, 484, L54, \textit{Paper\,I} 

\bibitem[Bellini et al.(2013)]{2013ApJ...769L..32B} Bellini, A., Anderson, J., Salaris, M., et al.\ 2013, ApJL, 769, L32 
    
\bibitem[Bellini et al.(2017)]{2017ApJ...844..164B} Bellini, A., Milone, A.~P., Anderson, J., et al.\ 2017, ApJ, 844, 164
  
\bibitem[Bellini et al.(2018)]{2018ApJ...853...86B} Bellini, A., Libralato, M., Bedin, L.~R., et al.\ 2018, ApJ, 853, 86

\bibitem[Campos et al. (2016)]{2016MNRAS.456.3729C} Campos, F., Bergeron, P., Romero, A. D., Kepler, S. O., et al. 2016, MNRAS, 456, 3729

\bibitem[Cassisi et al.(2014)]{2014A&A...571A..81C} Cassisi, S., Salaris, M., Pietrinferni, A., Vink, J.~S., \& Monelli, M.\ 2014, A\&A, 571, A81

\bibitem[da Silva et al.(2018)]{} da Silva, P., Steiner, J. E. and Menezes, R. B. 2018, ApJ, 861, 83

\bibitem[Gaia Collaboration et al.(2018)]{2018A&A...616A...1G} Gaia  Collaboration, Brown, A.~G.~A., Vallenari, A., et al.\ 2018, A\&A, 616, A1

\bibitem[Graham \& Driver(2005)]{2005PASA...22..118G} Graham, A.~W., \& Driver, S.~P.\ 2005, PASA, 22, 118

\bibitem[Gratton et al.(2005)]{2005A&A...440..901G} Gratton, R.~G., Bragaglia, A., Carretta, E., et al.\ 2005, A\&A, 440, 901 

\bibitem[Horowitz et al.(2010)]{2010PhRvL.104w1101H} Horowitz, C. J., Schneider, A. S., \& Berry, D. K.\ 2010, PhRvL, 104, 231101
  
\bibitem[Kalirai et al.(2009)]{2009ApJ...705..408K} Kalirai, J.~S., Saul Davis, D., Richer, H.~B., et al.\ 2009, ApJ, 705, 408 

\bibitem[Kalirai et al.(2012)]{2012AJ....143...11K} Kalirai, J.~S., Richer, H.~B., Anderson, J., et al.\ 2012, AJ, 143, 11

\bibitem[King et al.(2012)]{2012AJ....144....5K} King, I.~R., Bedin, L.~R., Cassisi, S., et al.\ 2012, AJ, 144, 5 

\bibitem[Libralato et al.(2018)]{2018ApJ...854...45L} Libralato, M., Bellini, A., Bedin, L.~R., et al.\ 2018, ApJ, 854, 45

\bibitem[Lindegren et al.(2018)]{2018A&A...616A...2L} Lindegren, L., Hern{\'a}ndez, J., Bombrun, A., et al.\ 2018, A\&A, 616, A2 

\bibitem[Milone et al.(2010)]{2010ApJ...709.1183M} Milone, A.~P., Piotto, G., King, I.~R., et al.\ 2010, ApJ, 709, 1183 

\bibitem[Milone et al.(2013)]{2013ApJ...767..120M} Milone, A.~P., Marino, A.~F., Piotto, G., et al.\ 2013, ApJ, 767, 120 

\bibitem[Milone et al.(2017)]{2017MNRAS.469..800M} Milone, A.~P., Marino, A.~F., Bedin, L.~R., et al.\ 2017, MNRAS, 469, 800

\bibitem[Milone et al.(2018)]{2018MNRAS.481.5098M} Milone, A.~P., Marino, A.~F., Renzini, A., et al.\ 2018, MNRAS, 481, 5098 

\bibitem[Milone et al.(2019)]{2019MNRAS.484.4046M} Milone, A.~P., Marino, A.~F., Bedin, L.~R., et al.\ 2019, MNRAS, 484, 4046, \textit{Paper\,II} 

\bibitem[Nardiello et al.(2015)]{2015A&A...573A..70N} Nardiello, D., Milone, A.~P., Piotto, G., et al.\ 2015, A\&A, 573, A70 

\bibitem[Nardiello et al.(2018)]{2018MNRAS.481.3382N} Nardiello, D., Libralato, M., Piotto, G., et al.\ 2018, MNRAS, 481, 3382 
  
\bibitem[Pietrinferni et al.(2006)]{2006ApJ...642..797P} Pietrinferni, A., Cassisi, S., Salaris, M., \& Castelli, F.\ 2006, ApJ, 642, 797 

\bibitem[Richer et al.(2013)]{2013ApJ...778..104R} Richer, H.~B. et al. 2013, ApJ, 778, 104

\bibitem[Robin et al.(2003)]{2003A&A...409..523R} Robin, A. C.,  Reyl\'e, C., Derri\`ere, S. \& Picaud, S. 2003, A\&A, 409, 523
  
\bibitem[Salaris et al.(2009)]{2009ApJ...692.1013S} Salaris, M., Serenelli, A., Weiss, A., \& Miller Bertolami, M.\ 2009, ApJ, 692, 1013 
  
\bibitem[Salaris et al.(2010)]{2010ApJ...716.1241S} Salaris, M., Cassisi, S., Pietrinferni, A., Kowalski, P.~M., \& Isern, J.\ 2010, ApJ, 716, 1241

\bibitem[Segretain \& Chabrier(1993)]{1993A&A...271L..13S} Segretain, L. \& Chabrier, G.\ 1993, A\&A, 271, L13 
   
\bibitem[Sarajedini et al.(2007)]{2007AJ....133.1658S} Sarajedini, A., Bedin, L.~R., Chaboyer, B., et al.\ 2007, AJ, 133, 1658 

\bibitem[Vandenberg et al.(2002]{2002ApJ...571..487V} VandenBerg, D.~A., Richard, O., Michaud, G., \& Richer, J.\ 2002, ApJ, 571, 487


\end{thebibliography}
\end{document}